\documentclass{iopart}

\usepackage{graphicx,iopams}  
\begin{document}

\title{STEP and fundamental physics}

\author{James Overduin,$^{1,2}$ Francis Everitt,$^3$ Paul Worden$^3$ and John Mester$^{3,4}$}

\address{$^1$ Department of Physics, Astronomy and Geosciences, Towson 
   University, Towson, MD 21252, USA}
\address{$^2$ Department of Physics and Astronomy, Johns Hopkins University,
   Baltimore, MD 21218, USA}
\address{$^3$ Hansen Experimental Physics Laboratory, Stanford University,
   Stanford, CA 94305, USA}
\address{$^4$ Associated Universities, Inc., Washington, DC 20036, USA}
\ead{francis@relgyro.stanford.edu}
\begin{abstract}
The Satellite Test of the Equivalence Principle (STEP) will advance 
experimental limits on violations of Einstein's Equivalence Principle
from their present sensitivity of 2 parts in $10^{13}$ to 1 part in $10^{18}$
through multiple comparison of the motions of four pairs of test masses of
different compositions in a drag-free earth-orbiting satellite.
We describe the experiment, its current status, and its potential 
implications for fundamental physics.  Equivalence is at the heart of
general relativity, our governing theory of gravity, and violations are
expected in most attempts to unify this theory with the other fundamental
interactions of physics, as well as in many theoretical explanations for
the phenomenon of dark energy in cosmology.  Detection of such a violation
would be equivalent to the discovery of a new force of nature.  A null result
would be almost as profound, pushing upper limits on any coupling between
standard-model fields and the new light degrees of freedom generically
predicted by these theories down to unnaturally small levels.
\end{abstract}

\maketitle

\section{Introduction}

STEP 
will perform a new test in Earth orbit of the Equivalence Principle, the foundation of our reigning theory of gravitation, general relativity (GR).\footnote{We refer throughout this article to the ``weak'' version of the EP, which states that gravitational and inertial mass are locally equivalent.}
There are good reasons, from most attempts to unify GR with the other fundamental interactions, and from recent developments in cosmology, for believing that at levels below the present limits of testing this 
principle may break down. 


STEP's sensitivity, five or six orders of magnitude better than any previous measurement, comes from a combination of the orbital environment and technological innovation. STEP takes  advantage of key technologies successfully developed at Stanford for the NASA Gravity Probe B mission (flight qualified SQUIDs, test mass caging mechanisms, drag-free control algorithms, proportional helium thrusters, charge measurement and UV discharge mechanism, etc.) as well as closely related cryogenic and other technologies.  Experience with the operation and data analysis for GP-B gives us great confidence in the projected sensitivity and has also shown both how to improve the measurement and how to reduce several disturbances. 

STEP will advance testing of the equivalence principle from $~2\times10^{-13}$ to $10^{-18}$. Violations in this range are associated with couplings between standard-model (SM) fields and the new light degrees of freedom generically predicted by leading attempts at unified theories of fundamental interactions (e.g., string theory) and cosmological theories involving dynamical dark energy (e.g., dilatons, moduli, quintessence). A null result would force existing constraints on these couplings down to unnatural levels, indicating that new approaches to unification and/or dark energy may be required. A detection would be equivalent to the discovery of a new force of nature. These are compelling reasons to include STEP along with particle accelerators (LHC) and orbiting telescopes (JDEM) as part of a concerted strategy to solve the deepest outstanding problems in physics.  But in the end, David Gross has remarked \cite{gro07}, ``any experiment that can advance limits on so fundamental a postulate by so many orders of magnitude justifies itself.''

\section{Theoretical motivation}

At the most basic level, a violation of the Equivalence Principle (EP) arises when a new field with gravitational-strength coupling is introduced into the Standard Model of particle physics (SM). Since this new field, unlike the metric or spin-two graviton field of general relativity, does not couple universally to SM fields, it introduces differences in their rates of fall. The simplest new fields usually considered are scalars; these feature generically in many attempts to account for the phenomenon of dark energy in cosmology, and in essentially all current approaches to the problem of unifying gravity with the other fundamental interactions of nature. Well-known examples whose implications for EP experiments have been explored in some detail include string-inspired dilaton and moduli fields \cite{dp94,ant01,dam02}, dynamical dark energy or quintessence \cite{car98,che05} and fields associated with variable fundamental coupling “constants” \cite{dz02,wet03}. More recent theoretical proposals with concrete and experimentally interesting predictions for EP violation include “chameleon fields” with density-dependent masses \cite{kw04,ms07}, modified-gravity or f(R) theories in which the new scalar is related to spacetime curvature \cite{ct08}, cosmological models in which a new field couples indirectly to SM fields via their interactions with dark matter \cite{car10} and perhaps most importantly, because of their generality, the various Lorentz-violating fields introduced as part of the Standard-Model Extension or SME \cite{kt09,kt10}.

Up to the present time, experimental efforts to detect the new fields predicted by unified-field theories and dark-energy models have depended on two main strategies. High-energy particle physicists make use of accelerators to look for missing-energy signatures that might be associated with the quanta of these fields (e.g. 
\cite{bar11}). Cosmologists employ large orbiting telescopes to look for changes in the bulk equation of state that might be associated with these fields as they evolve in time (e.g., 
\cite{geh10}). Space tests of the EP offer the possibility of a third, gravitational component in this search effort, one that that is complementary to the other two kinds of tests in that it holds the potential not only to demonstrate that a new field exists, but also to reveal how it interacts with the rest of the world \cite{ove08}.

Two arguments have historically been raised against the potential of EP tests to discover new fields with significant SM couplings. First, detectable EP violations are associated with long-range, and therefore light scalars; and there are good reasons to believe that new scalars in many of these theories should automatically acquire large masses. On the other hand, heavy scalars come with their own difficulties (e.g. the Polonyi or ``moduli problem,'' 
and light scalars appear more natural in the context of models for dark energy \cite{car98}. The latest studies derive new predictions for EP violation in an experimentally interesting range with light scalars whose SM couplings are assumed from the outset to be {\em weaker\/} than gravitational strength \cite{dd10}.

Second, allowing for the possibility that such fields might exist in principle, there is a belief in some quarters that they are of little interest in practice because their existence is already ruled out by existing experimental limits on EP violation. 
We wish to explore this argument carefully.
Violations can be constrained in many ingenious ways (see e.g. \cite{dim08,ni11}), but the most robust and sensitive experimental limits to date come from two methods: the use of moons and planets as test bodies in the gravitational field of another solar-system object (the ``celestial method,'' originally due to Newton) and the use of torsion balances (pioneered by E\"otv\"os) to probe the rates of fall of laboratory test masses in a horizontal component of the  gravitational field of the Earth, Sun or Milky Way. Current experimental upper limits from both approaches are similar. Lunar laser ranging constrains any difference in relative acceleration of the Moon and Earth toward the Sun to less than $(-1.0 \pm 1.4) \times 10^{-13}$ \cite{wil04}. State-of-the-art torsion-balance experiments imply that the relative rates of fall of beryllium and titanium toward the Earth differ by no more than $(0.3 \pm 1.8) \times 10^{-13}$ \cite{sch08}. Both techniques are pushing against fundamental limits and may not see major advances beyond the $10^{-13}$ level in the forseeable future.

To understand the theoretical significance of this number, and compare it to the planned STEP sensitivity of $10^{-18}$, it is helpful to focus on the simplest and most generic possible extension of the standard model with EP violation: one incorporating a single new scalar field $\phi$. Absent some protective symmetry (whose existence would require explanation), this new field couples to SM fields via dimensionless coupling constants $\beta_k$ (one for each SM field) with values not too far from unity. Calculations within the standard model (modified only to incorporate $\phi$) quantify the relationship between these constants and the relative rates of fall for test bodies with different elementary-particle compositions \cite{car98,kw00}. To order of magnitude, they show that the largest and most important of these couplings is the one associated with gluons, the gauge fields of quantum chromodynamics or QCD. 
Existing experimental upper limits of order $10^{-13}$ on the relative rate of fall $\Delta a/a$ translate into a constraint on this dimensionless QCD coupling constant of order $\beta_g < 10^{-6}$ \cite{che05}. From a particle-physics perspective, {\em it is this number\/} (and not $10^{-13}$) that provides the best gauge of the strength of existing constraints on violations of the EP.

Whether or not an upper limit like $10^{-6}$ proves that there is no point in looking further for EP-violating couplings may be debatable. To attach a measure of objectivity to this question, a short excursion into standard QCD proves instructive. There is a dimensionless coupling parameter $\theta$ in the Lagrangian of this theory, entirely analogous to the $\beta_k$, that one would naively expect to be of order one, but that must be less than $10^{-9}$ in actuality since any larger value would violate experimental upper limits on the electron dipole moment of the neutron \cite{din00}. 
Particle physicists regard this ``strong CP problem'' as so unnatural that they invoke a new particle to drive the offending parameter dynamically toward zero. That particle (the axion) is now one of two leading contenders for dark matter, and experiments are underway around the world to look for it.
Now return to the analogous ``EP problem.''  Existing upper limits on $\beta_k$ are of order $10^{-6}$. A null result at the STEP sensitivity would push this down to $10^{-9}$ \cite{ove08}. While no such argument can be entirely decisive, the smaller number closes the door on EP violation in a way that the larger one does not.  It forces theorists to the conclusion that the postulated EP-violating fields either do not exist, or that their SM couplings are suppressed by some new symmetry (analogous to the Peccei-Quinn or PQ symmetry associated with the QCD axion) whose existence would be as significant as that of the original field itself. 

We stress that this line of thinking does not predict that EP violations will occur at levels between $10^{-18}$ and $10^{-13}$ (though it is worth noting this range is singled out in most of the references cited above).  Rather, it says that violations {\em below\/} this range are so unnatural as to be not worth looking for.  In the words of Ed Witten \cite{wit00}, ``it would be surprising if $\phi$ exists and would not be detected in an experiment that improves bounds on EP violations by 6~orders of magnitude.''

The significance of this argument is that it makes STEP a ``win-win'' proposition, regardless of whether violations are actually detected. A null result would be almost as profound as a detection, because it would imply either that the new degrees of freedom generically predicted by unified and dark-energy theories either do not exist, or that there is some deep new symmetry or other mechanism that prevents their being coupled to SM fields. A historical parallel to such a null result might be found in the Michelson-Morley experiment, which reshaped physics precisely because it found nothing.  The ``nothing'' in that case finally forced physicists to accept the fundamentally different nature of light, at the cost of a radical revision of their concepts of space and time.  Non-detection of EP violation at the $10^{-18}$ level would suggest that gravity is so fundamentally different from the other forces that a similarly radical rethinking will be necessary to accommodate it within the same theoretical framework as the SM based on quantum field theory.

\begin{figure}[t!]
\begin{center}
\includegraphics[width=10.6cm]{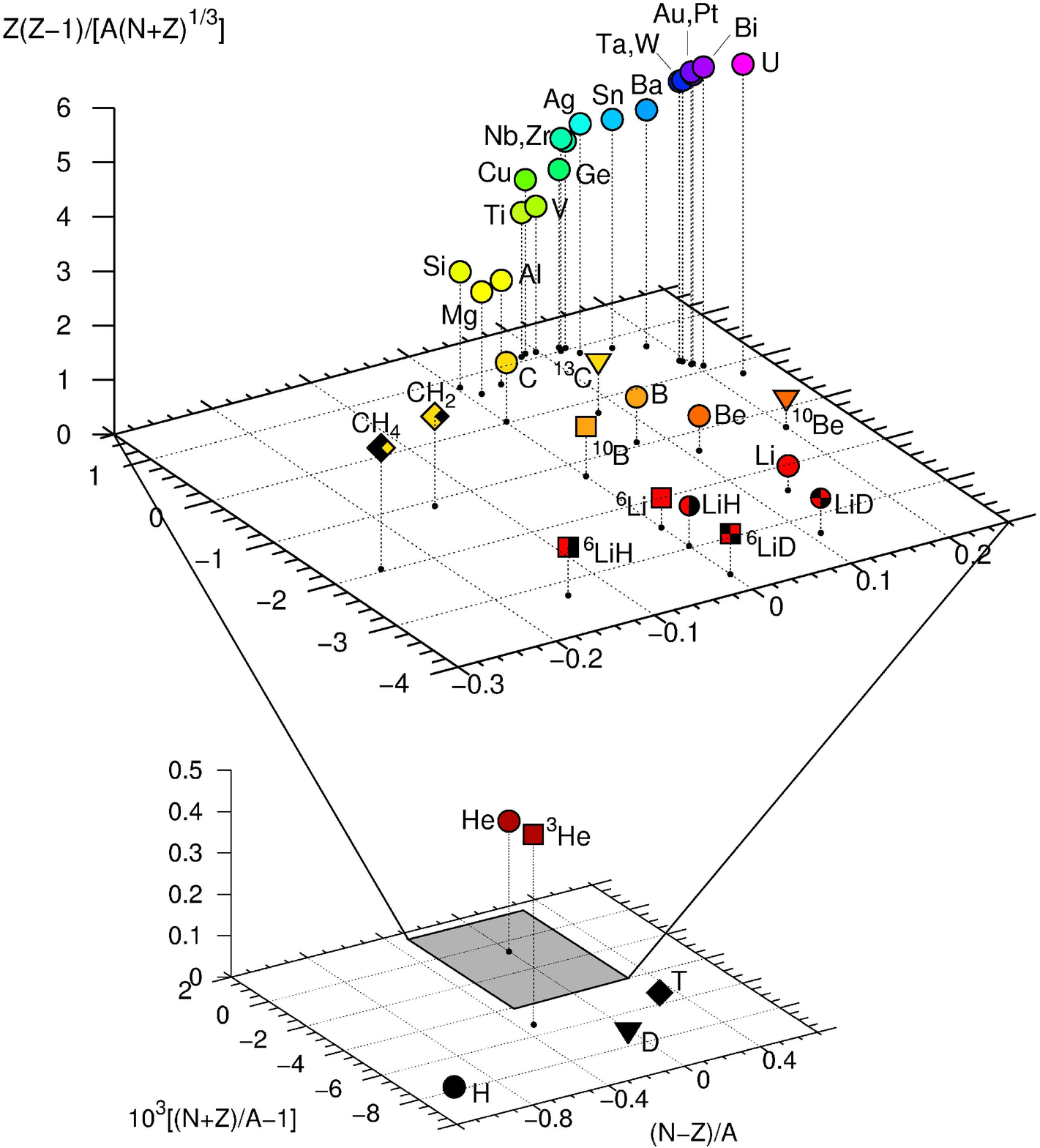}
\caption{Various elements and compounds in the phase space defined by
baryon number $N+Z$, neutron excess $N-Z$ and nuclear electrostatic energy
$[\propto Z(Z-1)]$, all normalized by atomic mass $A$.  The most common
isotopes are indicated with circles; lighter isotopes are squares; heavier
isotopes are triangles and diamonds; compounds are patterned.}
\label{fig-materials}
\end{center}
\end{figure}

Assuming that new EP-violating field(s) do exist, the choice of test-mass materials is important in maximizing the chance that they will be detected. Without knowing the nature of these fields in advance, it is impossible to anticipate every possibility, but general arguments like that that based on the minimal scalar field above are suggestive. The current STEP baseline (calling for four pairs of Pt-Ir, Nb and Be test masses arranged in a cyclic condition) is based on extensive theoretical work \cite{db94,dam96,bla01} showing that EP-violating fields in a general class of string-inspired models will couple most strongly to three kinds of ``elementary charge'': baryon number, neutron excess and electromagnetic self-energy. Recent developments suggest that strong nuclear binding energy may be equally important \cite{dd10,den08}. The general thrust of these studies is toward a configuration of test materials that spans the largest possible volume in the phase space of elementary charges (Fig.~\ref{fig-materials}). This goal must of course be balanced against practical considerations such as manufacturability, stability, cost and the need to ensure that any effect seen is real and robust (see discussion below).
The use of multiple pairs of broadly different test materials allows STEP to probe the largest possible volume of this phase space so that any violation, once seen, can be used to discriminate between competing theories. This is a key strength of the mission relative to other proposed experiments employing as test bodies isotopes of the same element \cite{dim08}\footnote{Atom interferometric EP tests involving different atoms, such as lithium and cesium \cite{km10}, would be an important advance in this regard.} or a maximum of two different materials \cite{tou09}, and is arguably as important as the improvement in sensitivity to EP violation itself.

\section{Science concept and design motivation}

\begin{figure}[t!]
\begin{center}
\includegraphics[width=5.7cm]{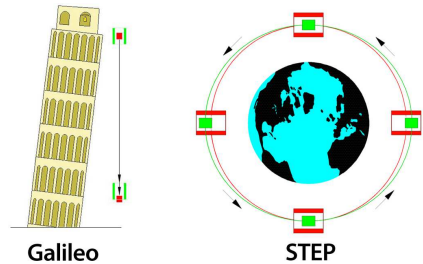}
\caption{STEP concept: Galileo's free-fall experiment in orbit}
\label{fig-concept}
\end{center}
\end{figure}

The simplest conception of STEP is a Galileo free-fall experiment in Earth orbit: independent masses of different composition fall toward the Earth, but in orbit the experiment  never stops because they do not impact the ground
(Fig.~\ref{fig-concept}).
The signal is driven by almost the full gravity of the earth, and the measurement is easier because the instrument falls along with the masses.  The masses are subjected to fewer disturbances because of the high vacuum of space.  This  simplistic concept must be modified  by orbital dynamics and unavoidable (or even necessary) couplings between the masses and the measuring instrument, as well as many practical considerations. We have tried to keep as close to the ideal as possible.  This approach assures that the necessary differences from an ideal measurement result in minimal error in the measurement.  Thus, for example, it is important to shift the frequency of the EP signal away from the orbit frequency of about $1.9 \times 10^{-4}$~Hz because several disturbances (e.g. air drag, solar pressure) are expected to occur in synchrony with the orbit, and accordingly during the mission we plan to rotate the experiment at several rates in the range $\pm 3$ times orbit frequency during the mission. To the extent that the rotation is constant, it cannot introduce disturbances at other than at DC, and the only change to the basic concept is that the masses and sensor rotate as they fall, modulating the signal from any EP violation. Multiple rates help distinguish the signal from other disturbances (e.g., solar heating, helium tides).

A violation of the EP causes the masses to fall at different rates.  An EP measurement is just a time sequence of acceleration measurements made under specified conditions, which can be analyzed to distinguish possible violations of EP from various disturbances.  STEP's primary data comprises the common and differential mode accelerations of four pairs of test masses of different materials.  The measurements must be of adequate sensitivity to achieve the science goal of one part in $10^{18}$, and free of disturbance to approximately the same level.  It is vitally important to be able to show that what is measured is not just noise or a disturbance. 

Thus, quality of data is more important than quantity. Quantity of data allows averaging over stochastic and some regular disturbances, but does not increase scientific certainty in the presence of systematic errors.  To show that any small difference in acceleration is not caused by an unknown disturbance, all relevant experimental conditions (e.g. spacecraft configuration and status, external magnetic field, particle radiation) need to be measured.  Disturbances to the test masses must be demonstrably smaller than the desired sensitivity of $10^{-18}$, or easily distinguishable from the signal of an EP violation. 

In such a sensitive measurement is necessary to keep all disturbances small: ideally, zero or much smaller than the quantity to be measured.  This is consistent with the ideal of completely freely falling masses, but in practice, large disturbances (even though they do not resemble the signal) are easily converted into the signal bandwidth by nonlinear coupling. Ordinary thermal noise can be minimized by operating at cryogenic temperature, which also vastly reduces gas pressure, thermal expansion and stability issues, and many other problems. But the largest source of disturbances is the spacecraft itself, especially those portions of the instrument which are physically closest to the test masses. Special care must be taken in those areas.  

Motion of the instrument/spacecraft relative to the masses must be kept to a minimum because large motions couple directly into the measurement, and because unavoidable small forces (e.g. magnetic, electrostatic, and gravitational) between the spacecraft and mass will result in disturbance by nonlinear coupling to the masses.  Drag-free control, in which the spacecraft is controlled to follow the test masses, is an excellent solution as demonstrated by GP-B. 

\begin{figure}[t!]
\begin{center}
\includegraphics[width=12.3cm]{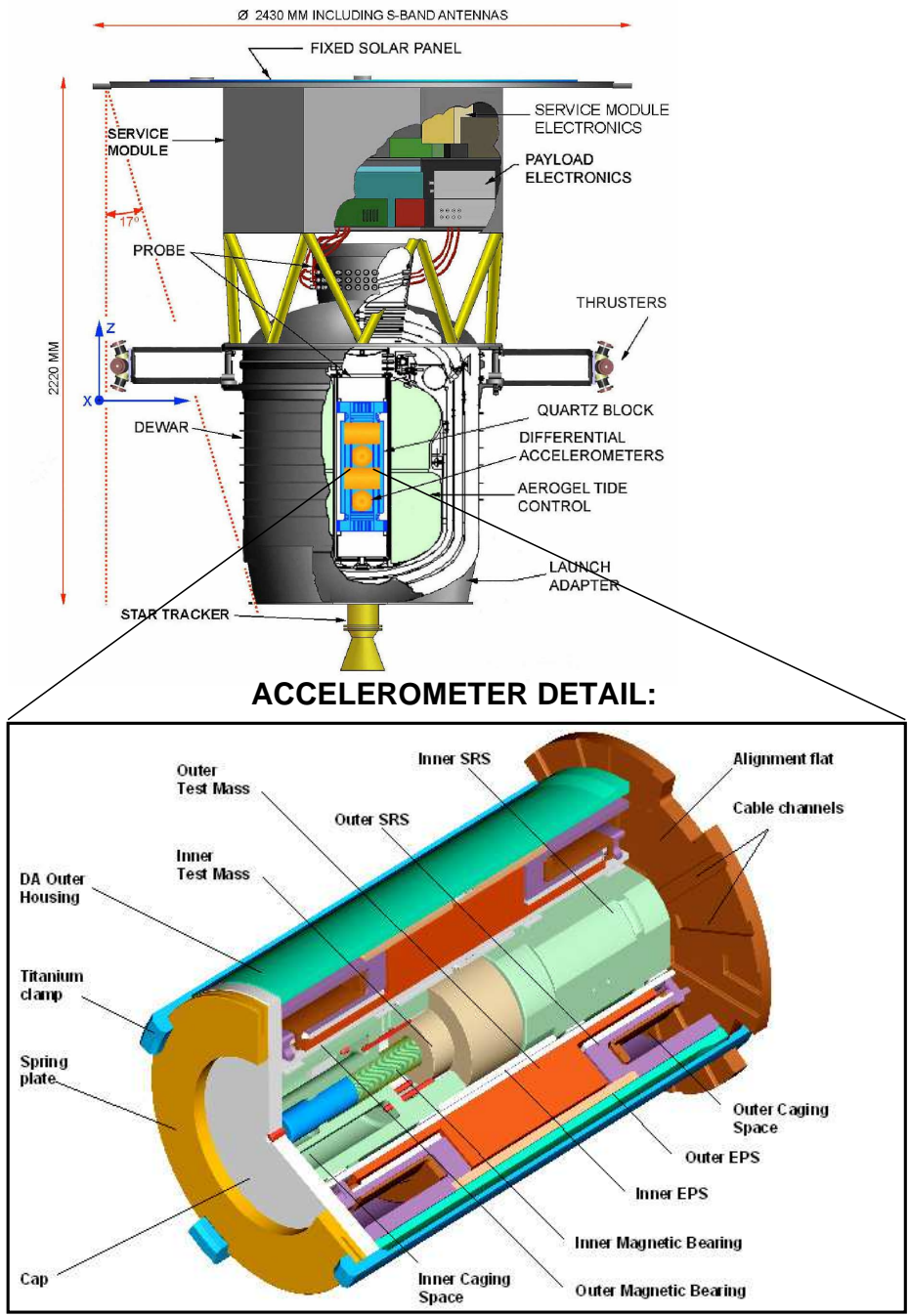}
\caption{Schematic cutaway of the STEP spacecraft (top) with accelerometer detail (bottom; DA=differential accelerometer, EPS=electrostatic positioning system, SRS=squid readout sensor).}
\label{fig-schematic}
\end{center}
\end{figure}

The measurement system cannot be perfect and will have some sensitivity to modes other than the differential mode which contains the EP signal.  To address this issue we constrain the masses' motion to one dimension, with the other degrees of symmetry controlled by superconducting magnetic bearings. The bearings provide stiff radial constraint with less than $10^{-6}$ of the radial force appearing in the axial direction.  This suggests a cylindrical geometry for the accelerometers and test masses (Fig.~\ref{fig-schematic}).  It is then easy to use  test masses centered on each other to reduce forces from the earth's gravity gradient. 

Measurement of the test masses' accelerations inevitably disturbs them.  STEP has chosen SQUID sensors because of their inherent sensitivity and stability; the magnetic fields from the sensing coils produce a spring constant tending to restore the masses to a central position. The resulting system is an accelerometer rather than a position sensor. It is optimized not by great position sensitivity but by the ratio of displacement caused by acceleration to the smallest measureable distance. The displacement decreases much faster than the measurable distance, so we operate with very weak springs ($\sim\!1000$~s period) and relatively poor position sensitivity ($\sim\!10^{-13}$~m). 

Much of STEP's observation time is committed to repeating measurements with deliberate changes in experimental conditions to distinguish and quantify systematic errors in the signal. Such qualitative, interactive changes distinguish laboratory science from observational science. We will vary parameters such as temperature, test mass charge, or satellite roll to assess their influence as systematic effects. This disturbance assessment is critical to the credibility of the experiment.

The science data quality can be greatly enhanced by simultaneous measurements. Operating several accelerometers at the same time gives independent measurements under almost identical conditions.  Comparing the resulting time sequences quickly identifies which instruments are responding differently, regardless of the cause.  Using simultaneous measurements with three or more materials in a ``cyclic condition'' we can verify or rule out systematic disturbances. That is, given three accelerometers comparing materials A,B,C, the sum of differential acceleration measurements A-B, B-C, and C-A should be zero. If not, there is certainly a systematic error.

Simultaneous operation requires proper orientation of the satellite, to prevent overlarge disturbances caused by the earth's gravity gradient acting between accelerometers. Within any one accelerometer, the test masses need to have nearly coincident centers of mass for the same reason.  Fortunately the gravity gradient forces occur at twice the frequency of the EP signal, and are proportional to the distance between the centers of mass.  They can be used as an error signal to move the mass centers into coincidence.  They also give a method for measuring the satellite's orientation relative to the earth's gravity gradients and reorienting the satellite accordingly.  Since the gravitational disturbance is its own error signal, it can be eliminated almost completely, at all frequencies.  Centering the masses on each other reduces their differential response to all gravity gradients, and orienting the satellite so that the accelerometers lie on a horizontal line minimizes the total gravitational disturbance.  By rotating the satellite, we separate the EP signal from periodic disturbances known or suspected to exist in orbit: this requires that the rotation be about, and the accelerometers should lie on, the pitch axis. 

An additional improvement is accomplished by optimizing the test mass shape.  Ideal spherical masses centered one inside the other would eliminate the gravitational disturbance from residual motion of the nearby satellite and liquid helium, but provide no access enabling measurement of the inner test mass.  STEP's masses are cylindrical, with equal moments of inertia along each axis. This acts like a point mass, so far as uniform gravity gradients are concerned.  Similar optimizations adding ``belts'' of material around the test masses result in masses which respond identically to fourth order in the gravity gradient, reducing the problem  to insignificance. 

Other disturbances are less under the control of the experimenter but can be accounted for by measurement and/or error analysis.  For example, time varying electrical charge on the test mass might cause a force between the test mass and its housing, mimicking an EP violation.  The test masses (rotors) in GP-B showed no significant change in charge at the level of STEP, and STEP carries a similar measurement and UV discharge system. We therefore expect to be able to measure and model electrical forces to the level required.  Patch effect forces, in contrast, were a serious disturbance on GP-B, but they enter STEP in a very different way which is included in our error analysis \cite{wm09,fer10}.

\section{Mission design and technical implementation}

STEP has evolved through a series of studies and technology development programs sponsored in the U.S.A. by NASA and in Europe by ESA and European national agencies.  The original concept originated in 1972 and a ground-based version was built and operated at Stanford University with NASA and NSF support.  In 1989, with NASA Code~U backing, STEP was proposed to the ESA~M2~AO as a joint U.S./European mission. It was one of four from across the whole range of ESA science to be awarded a Phase~A Study, and ranked second in the final 1993 M2 selection. 

From 1992 NASA Code~U provided enhanced instrument development funding.  In 1998, STEP successfully passed the dual Science Concept Review (SCR), Requirements Definition Review (RDR) selection process used to determine whether a Code~U program should enter the queue for flight.  In 1999, STEP was selected for Phase~A Study under SMEX 8/9. This Phase~A study, completed in 2002, was conducted in collaboration with JPL and defined much of what is now the current baseline design. In parallel with this NASA Phase~A study, ESA sponsored two industrial studies of the STEP spacecraft---the so-called Service Module---carried out by ASTRIUM UK and a European payload feasibility study conducted by collaborating institutions in France, Germany, Italy, the Netherlands and the UK.  

In 2004 STEP defined a technology development phase under the sponsorship of NASA MSFC. This program resulted in the fabrication and test of an engineering unit inner accelerometer.  A further Service Module proposal was conducted jointly with ASTRIUM and Surrey Satellite Limited in 2008. 

As stated above, STEP mission design is based around 4 differential accelerometers.  These will be supported by a quartz block assembly housed in a 220~liter superfluid helium dewar.  Spacecraft mass and power specifications are 700~kg and 300~W respectively.  A 550~km high, sun-synchronous orbit (97$^{\circ}$ inclination) mitigates thermal disturbances over the 6~month mission lifetime.  The spacecraft is slowly rolled to shift the science (EP violation) signal frequency away from the orbit frequency.  Mechanical disturbances are reduced by a drag-free control system that directs the helium boil-off gas through an array of proportional thrusters.  

Cryogenic operation of the instrument provides many advantages including greatly improved thermal and mechanical stability, greatly reduced thermal expansion, ultra-low gas pressure, superconducting shielding reducing external magnetic field variations to $10^{-11}$ of their external value, reduced thermal noise, and the ability to use extremely sensitive SQUID position readouts. The accelerometers will be housed in a vacuum probe inside a cryogenic dewar. 
The baseline dewar is based on an existing flight-qualified vessel 1.3~m long and 1.2~m in diameter, built by Lockheed Martin. The main tank contains 200~liters of superfluid helium at an operating temperature of 1.8K. It is supported from the vacuum shell by six passive orbital disconnect struts (PODS-V). There is also a normal helium guard tank. Cryogenic temperature is maintained in the instrument by continuously boiling helium liquid as a coolant. The surplus gas, which would otherwise cause a disturbance, is used by the drag-free system as propellant. The on-orbit lifetime is calculated to be 8~months.  Helium slosh and tide disturbances are mitigated by filling the helium tank with aerogel. Ground tests have verified that the aerogel system survives launch loads and vibrations.
 
A Service Module (SM), which provides standard power, communications, command sequencing and safe modes is based on ESA-sponsored STEP studies conducted by ASTRIUM UK and Surrey Satellites Ltd.  Characteristics of the proposed SM design configuration include: (1)~no moving parts or matter in science measurement mode, (2)~thermal design using passive control methods only (no fluid movement), (3)~accommodation of payload electronics with minimum variation in temperature and (4)~a stiff fixed solar array structure mounted to an octagon and strut structure with no frequency coupling into the experiment that impacts science data.
The SM structural elements, octagonal walls, payload pallet, and struts will be made from CFRP in order to minimize the thermal and mechanical distortions that could affect the EP measurement. The fundamental frequencies are 46~Hz axial and 26~Hz lateral.  A structure test model of the octagon and floor has completed qualification testing to levels well beyond those needed for STEP. 

The power subsystem will control the bus voltage with solid-state power controllers that switch the solar array strings. This approach can provide well-regulated bus voltage to the payload since there will be no eclipses during the mission. A Gallium Arsenide solar array will provide ample power margin for ASH (acquisition and safe hold) and Science Mode.  The power budget will be nearly constant throughout the entire mission. The fixed array/sun shield which is not fully populated is sized to accommodate an off-Sun angle of up to 20$^{\circ}$.
SM avionics architecture will be based on the centralized On-Board Management Unit (OBMU) developed and qualified in the framework of the European small satellite programs.   

Complete $4\pi$-steradian communications coverage will be obtained by mounting two antennas at opposite positions on the solar array.   With a 5~W transmitter feeding both antennas simultaneously, a data rate of 1.6~Mbps will be achieved with a downlink margin of at least 9~dB. With off-the-shelf redundant transceivers in a single unit for communication station at Svalbard, 60~hours of stored data could be downlinked in 8~orbits. 
The data handling system will provide SM command and control, manage the provision of attitude and navigation sensor data to the payload, and store payload data. The OBMU will control all command and data handling and all of the spacecraft system interfaces. A built-in 4.3~Gb solid-state Mass Memory will be used to store collected science data between visibility periods, providing 60~hours of storage at 20~kbps. The link between the SM and the PLM will be via a MIL-STD-1553 bus. The SM computer will serve as the 1553 master throughout all phases of the mission. 

The STEP thermal design takes advantage of benign orbit conditions, using passive thermal control techniques to minimize thermal distortion and heat flow to the dewar, and to provide stable temperatures for the readout electronics. The slow roll and dawn/dusk orbit produce almost uniform temperature distribution around the spacecraft. Multilayer insulation will surround structures (struts, dewar shell) to minimize heat load from albedo and the resulting temperature variations. The solar array will shade the outer shell of the dewar to minimize its operating temperature. Heat transfer from the SM to the dewar will be minimized by conductive and radiative isolation, so that the dewar outer shell will be maintained at nominally 220~K throughout the mission. The payload electronics units will not be directly mounted to the radiating surface of the octagon, minimizing their orbital temperature variation.

Three launch vehicles are currently being considered. The Russian Rokot, marketed by Eurockot of Bremen, Germany, meets STEP launch requirements with 30\% weight margin. The Space~X Falcon~9 has recently flown and greatly exceeds STEP requirements. The Indian Polar Satellite Launch Vehicle (PSLV) has had a string of successes with Indian and international payloads. PSLV weight margin for STEP is slightly greater than 100\%.

STEP has nearly completed its technology development phase and is now in position to compete for a flight mission new start.
Phase~A study grass-roots costing, independently verified using the Aerospace Small Satellite Cost Model and the JPL Parametric Mission Cost Model, confirm that STEP is a SMeX (Small Explorer) class mission.  Opportunities to re-compete STEP within the NASA explorer program will be enhanced following the recommendation of the  NRC Decadal Survey Report, {\em New Worlds New Horizons\/}, to augment the NASA Explorer program.

\ack J.O. is supported by a Towson University College of Graduate Studies and
Research fellowship.

\end{document}